\documentclass[aps,nofootinbib,commonaddress,preprint,showpacs]{revtex4}
\usepackage{amsmath}
\usepackage{graphicx}
\usepackage{color}

\begin{document}

\title{On kinetic description of electromagnetic processes in a quantum plasma}
\author{Yu.~Tyshetskiy}
\email{y.tyshetskiy@physics.usyd.edu.au}
\author{S.V.~Vladimirov}
\author{R.~Kompaneets}
\affiliation{School of Physics, The University of Sydney, NSW 2006, Australia}

\date{\today}
\received{}

\begin{abstract}
A nonlinear kinetic equation for nonrelativistic quantum plasma with electromagnetic interaction of particles is obtained in the Hartree's mean-field approximation. It is cast in a convenient form of Vlasov-Boltzmann-type equation with ``quantum interference integral'', that allows for relatively straightforward modification of existing classical Vlasov codes to incorporate quantum effects (quantum statistics and quantum interference of overlapping particles wave functions), without changing the bulk of the codes. Such modification (upgrade) of existing Vlasov codes may provide a direct and effective path to numerical simulations of nonlinear electrostatic and electromagnetic phenomena in quantum plasmas, especially of processes where kinetic effects are important (e.g., modulational interactions and stimulated scattering phenomena involving plasma modes at short wavelengths or high-order kinetic modes, dynamical screening and interaction of charges in quantum plasma, etc.) Moreover, numerical approaches involving such modified Vlasov codes would provide a useful basis for theoretical analyses of quantum plasmas, as quantum and classical effects can be easily separated there.
\end{abstract}

\maketitle


\newpage

\section{Introduction \label{sec:intro}}

The interest in quantum plasmas -- plasmas with quantum effects playing a significant role in their collective behavior -- has considerably increased in the recent decade, during which a significant number of publications appeared on this subject (see, e.g.,~\cite{Shukla_Eliasson_PPCF_2010,Shukla_Eliasson_UFN_2010,Stenflo_etal_EPL_2006,Melrose_Mushtaq_2009,Kuz'menkov_Maksimov_2002,Vlad_Tysh_UFN_2011,Vladimirov_PPCF_2011,Krivitskii_Vladimirov_1991,Eliasson_Shukla_JPP_2009,Marklund_etal_2008,Brodin_etal_2011,Bonitz_etal_2003} and references therein). This surge of interest can primarily be associated with the recent progress in manufacturing and manipulation of metallic and semiconductor nanostructures, whose properties are to a large extent governed by collective (plasma) effects of their electron (and hole) population. Quantum plasma effects may also become important in fast ignition scenario of modern inertial confinement fusion (ICF) experiments, when deuterium-tritium plasma is compressed to very high (super-solid) densities by intense laser pulses~\cite{Son&Fisch_2005,Lindl_1995,Tabak_etal_2005}. Recent spectral measurements of x-ray Thomson scattering~\cite{Kritcher_etal_2008,Lee_etal_2009} enable precise probing of electron distribution function in warm dense matter regime~\cite{Glenzer_etal_2007}, which opens the possibility of direct experimental studies of weakly and strongly coupled quantum plasmas. All these developments make quantum plasma regime both accessible and relevant, making theoretical efforts in describing quantum plasmas, and in particular linear and nonlinear collective processes in quantum plasmas, both timely and important. 

Microscopic theoretical models of quantum plasma usually treat the plasma as a weakly coupled gas of particles, described in the framework of the Hartree's mean-field approximation~\cite{Klim_Statfizika}. We should note that not all quantum plasmas are weakly coupled, e.g., for conductivity electrons in metals the potential energy of their interaction with each other and with the lattice ions is comparable to or exceeds their mean kinetic energy, thus making them a moderately coupled quantum liquid for which the mean-field approximation is, strictly speaking, invalid. However, even for such quantum plasmas the simple mean-field collisionless approximation yields a good qualitative picture of many collective phenomena such as waves and their interactions, while accounting for particle correlations (e.g., due to collisions and exchange interaction of particles) only yields quantitative corrections to the results of the mean-field theory; see, e.g., Ref.~\cite{Klim_Silin_UFN_1960}. For this reason, here for simplicity we restrict to the mean-field collisionless approximation for quantum plasmas. This assumption is quantitatively justified for plasmas with a small coupling parameter~\cite{Shukla_Eliasson_UFN_2010} $\Gamma\ll 1$, and captures most qualitative features of collective modes even for plasmas with $\Gamma\sim 1$, where $\Gamma=U_{\rm int}/\epsilon_{\rm kin}\sim e^2 n^{1/3}/\epsilon_{\rm kin}$, $U_{\rm int}$ and $\epsilon_{\rm kin}$ are the mean potential and kinetic energies of plasma particles, $e$ is the particle charge, $n$ is the particle number density.

The linear theory of weakly coupled quantum plasma, both unmagnetized and magnetized, has been developed by 1960s~\cite{Gol'dman_1947,Lindhard_1954,Klim_Silin_UFN_1960,Kelly_1964}. However, quantum plasmas are often formed or exist under the conditions of powerful energy input (e.g., in ICF experiments), leading to a rapid development of nonlinear phenomena, which require a proper theoretical treatment. A number of papers (see, e.g., Refs~\cite{Garcia_etal_2005,Brodin_Marklund_PoP_2007,Shukla_etal_2009,Shukla_Eliasson_UFN_2010,SSB_2010,Brodin_etal_2011} and references therein) appeared recently addressing various nonlinear phenomena in quantum plasmas by employing the so-called quantum fluid theory (QFT) approach~\cite{Manfredi_Haas_PRB_2001,Haas_PoP_2005,Haas_etal_NJP_2010,Brodin_Marklund_PRE_2007,Shukla_Eliasson_UFN_2010}. The QFT is considerably simpler than the more general kinetic theory (e.g., the theory based on the kinetic equation for the quantum plasma distribution function, the so-called Wigner function~\cite{Tatarskii_1983}), as QFT operates with quantities (fluid density, flow velocity, pressure) that depend on four variables $\mathbf{r},t$, while the kinetic theory operates with the quantum distribution function $f$ that depends on seven variables $\mathbf{r},\mathbf{p},t$, where $\mathbf{r}$ and $\mathbf{p}$ are position and momentum vectors, respectively, and $t$ is time. This relative simplicity makes the QFT description more favorable (compared to the kinetic description) when treating nonlinear phenomena, however, it comes at a cost of rather restricted validity range of such description and, correspondingly, its results (the characteristic length scale of waves or nonlinear structures should be large compared to the Thomas-Fermi length in degenerate quantum plasma, or to the Debye length in non-degenerate quantum plasma, for QFT to be valid). This restriction appears as a result of the assumptions necessarily made in the course of derivation of the ``collisionless QFT'' from the kinetic theory of collisionless quantum plasma~\cite{Manfredi_Haas_PRB_2001,Vlad_Tysh_UFN_2011}. Also, as all fluid theories, QFT does not account for essentially kinetic effects, such as, e.g., Landau damping~\cite{Landau_1946}. Because of its restricted validity, QFT is unable to correctly describe nonlinear wave interaction processes such as, for example, stimulated Raman scattering (SRS) and stimulated Brillouin scattering (SBS) of laser light off electrostatic plasma modes at short wavelengths (note that, unlike in classical plasmas, in degenerate quantum plasma electrostatic modes can have zero damping at short wavelengths~\cite{Krivitskii_Vladimirov_1991}, and thus can effectively participate in nonlinear interactions of electrostatic and electromagnetic waves), or other modulational-type interactions involving higher-order kinetic modes in quantum plasma. Hence, a kinetic model is needed for proper description of such processes. Proper description of dynamical screening and interaction of charged bodies in quantum plasma also requires a kinetic treatment~\cite{Else_etal_PRE_2010,Else_etal_EPL_2011,Vladimirov_PPCF_2011}.

In this paper, we develop, in the framework of Hartree's mean-field approximation~\cite{Klim_Statfizika}, a kinetic model of a nonrelativistic collisionless quantum plasma consisting of spinless charged particles with electromagnetic interaction. This model covers a wider range of plasma parameters than the classical kinetic model based on Vlasov equation coupled with Maxwell's equations, and should be useful for studying a wide range of linear and nonlinear electrostatic and electromagnetic processes in unmagnetized or magnetized uniform or nonuniform quantum plasmas, for which the effect of collisions, particle spin and relativistic effects are unimportant or only lead to minor quantitative corrections. (Hence this model is not expected to be applicable to, e.g., SRS of relativistically strong laser pulses in plasma, and is not capable of describing, e.g., spin waves). The main result is the kinetic equation for the one-particle quantum distribution function (one-particle Wigner's function $f_1$) of such system, which is presented in the form of a ``classical'' Vlasov-Boltzmann equation for $f_1$ (which by itself contains quantum effects), with explicit quantum terms appearing on the right-hand side in a ``quantum interference integral'' (manifesting the effect of quantum interference of overlapping particle wave functions, and vanishing in the classical limit). This form of the quantum kinetic equation allows for its rather straightforward implementation in existing numerical Vlasov codes, whose only modifications would be the inclusion of the ``quantum interference integral'' as a source term in the Vlasov equation, coupled with the correct initial and boundary conditions for the quantum distribution function (the latter in general can be different from those for the classical distribution function, see~\cite{Tatarskii_1983}). Existing classical Vlasov codes, modified in this way, can be used to simulate various linear and nonlinear collective effects in nonrelativistic quantum plasmas.

\section{Kinetic model of quantum plasma with electromagnetic interaction}
In this section, we start with recalling the well-known basics of the quantum statistical theory, and then proceed to obtain the kinetic equation for a quantum plasma with spinless particles interacting electromagnetically, which is the main result of this paper.
\subsection{Quantum distribution function}
Statistical description of a system of many interacting quantum particles (quantum plasma) is done in terms of its density matrix, which allows us to obtain the mean values and probability distributions of all the physical parameters of the system. In particular, it is convenient to describe the quantum plasma in terms of the density matrix in the mixed coordinate-momentum representation -- the quantum distribution function $f_N(\mathbf{r}_N,\mathbf{p}_N,t)$, suggested by Wigner~\cite{Wigner_1932} and sometimes called the $N$-particle Wigner function, defined as
\begin{equation}
f_{N}(\mathbf{r}_N,\mathbf{p}_N,t) = \frac{1}{(2\pi)^{3N}}\int d \vec{\lambda}\ {\rm e}^{-i \vec{\lambda}\cdot\mathbf{p}_N}
\rho_N(\mathbf{r}_N-\hbar\vec{\lambda}/2,\mathbf{r}_N+\hbar\vec{\lambda}/2,t),
\label{eq:f_N_def}
\end{equation}
where $\rho_N(\mathbf{r}_N,\mathbf{r}'_N,t)$ is the density matrix of the system in the coordinate representation, $N$ is the number of particles in the system, $\mathbf{r}_N$ and $\mathbf{p}_N$ are $3N$-dimensional vectors denoting the sets of coordinates and canonical momenta of all system particles, and $\hbar$ is the reduced Planck constant. In the classical limit $\hbar\rightarrow 0$, $f_N$ becomes the classical $N$-particle phase space distribution function, hence the description in terms of the Wigner function covers both classical and quantum plasmas. The properties of the Wigner function are discussed in detail by, e.g., Tatarskii~\cite{Tatarskii_1983}. We note that the Wigner function~(\ref{eq:f_N_def}) should not be thought of as the density of the system's states in $\mathbf{r}_N,\mathbf{p}_N$ phase space, as it is not positively definite (i.e., can attain negative values) due to non-commutativity of coordinate and momentum operators in quantum mechanics~\cite{Tatarskii_1983}.

Equation governing the evolution of $f_N$ is obtained from the evolution equation for the density matrix $\rho_N(\mathbf{r}_N,\mathbf{r'}_N,t)$ and reads~\cite{Moyal_1949,Klim_Silin_UFN_1960,Klim_Statfizika}
\begin{eqnarray}
\frac{\partial f_N(\mathbf{r}_N,\mathbf{p}_N,t)}{\partial t} = \frac{1}{(2\pi)^{6N}}\frac{i}{\hbar}\int\ldots\int d \vec{\lambda}_N
d\mathbf{k}_N d \vec{\eta}_N d\mathbf{q}_N\ {\rm e}^{i\left[ \vec{\lambda}_N\cdot( \vec{\eta}_N-\mathbf{p}_N)+\mathbf{k}_N\cdot(\mathbf{q}_N-\mathbf{r}_N)\right]}
f_N(\mathbf{q}_N, \vec{\eta}_N,t)
\nonumber\\
\times\left[\mathcal{H}\left(\mathbf{q}_N-\frac{1}{2}\hbar \vec{\lambda}_N,\vec{\eta}_N+\frac{1}{2}\hbar\mathbf{k}_N\right)
- \mathcal{H}\left(\mathbf{q}_N+\frac{1}{2}\hbar \vec{\lambda}_N,\vec{\eta}_N-\frac{1}{2}\hbar\mathbf{k}_N\right)\right],
\label{eq:f_N_kinetic}
\end{eqnarray}
where $\mathcal{H}(\mathbf{r}_N,\mathbf{p}_N,t)$ is the system's Hamiltonian function containing exact (not averaged) fields through which particles interact.

Description of plasma in terms of the $N$-particle quantum distribution function $f_N(\mathbf{r}_N,\mathbf{p}_N,t)$ is very cumbersome, as $f_N$ depends on a huge number of variables, $6N+1$. However, for most physical applications it is sufficient to know the one-particle quantum distribution function $f_1(\mathbf{r}_1,\mathbf{p}_1,t)$ that depends on coordinates $\mathbf{r}_1$ and momenta $\mathbf{p}_1$ of one particle, irrespective of positions and momenta of all other particles in the system:
\[
f_1(\mathbf{r}_1,\mathbf{p}_1,t)\equiv\int{f_N(\mathbf{r}_N,\mathbf{p}_N,t) d\mathbf{r}_2\ldots d\mathbf{r}_N d\mathbf{p}_2\ldots d\mathbf{p}_N}.
\]
The equation for $f_1$ can be obtained from~(\ref{eq:f_N_kinetic}) using the approach similar to the BBGKY (Bogoliubov-Born-Green-Kirkwood-Yvon) hierarchy approach for classical plasma~\cite{Balescu_1975,Klim_Statfizika}; the resulting equation for $f_1(\mathbf{r}_1,\mathbf{p}_1,t)$ contains the two-particle Wigner distribution function $f_2(\mathbf{r}_1,\mathbf{r}_2,\mathbf{p}_1,\mathbf{p}_2,t)$, the equation for which in turn contains the three-particle distribution function, and so on. The resulting set of coupled equations -- a quantum analogue of the BBGKY hierarchy -- is equivalent to Eq.~(\ref{eq:f_N_kinetic}), and its solution is equally difficult. However, in systems of weakly interacting particles, with $\Gamma=U_{\rm int}/\epsilon_{\rm int}\ll 1$, the problem of finding $f_1$ can be significantly simplified by neglecting the two-particle correlation function $g_2(\mathbf{r}_1,\mathbf{r}_2,\mathbf{p}_1,\mathbf{p}_2,t)$ defined by the following relation~\cite{Klim_Statfizika}:
\begin{equation}
f_2(\mathbf{r}_1,\mathbf{r}_2,\mathbf{p}_1,\mathbf{p}_2,t)=f_1(\mathbf{r}_1,\mathbf{p}_1,t)f_1(\mathbf{r}_2,\mathbf{p}_2,t)+g_2(\mathbf{r}_1,\mathbf{r}_2,\mathbf{p}_1,\mathbf{p}_2,t). \label{eq:g_2}
\end{equation}
The function $g_2$ characterizes statistical correlation between particles $1$ and $2$ due to their interaction. It defines the value of the collision integral $I(\mathbf{r}_1,\mathbf{p}_1,t)$ in the evolution equation for $f_1$, and hence neglecting $g_2$ in the case of weak particle interactions corresponds to the collisionless plasma approximation, $I(\mathbf{r}_1,\mathbf{p}_1,t)=0$. The resulting approximate equation for $f_1(\mathbf{r}_1,\mathbf{p}_1,t)$ does not contain $f_2$, having the form of Eq.~(\ref{eq:f_N_kinetic}) with $N=1$, in which the Hamiltonian function $\mathcal{H}(\mathbf{r}_1,\mathbf{p}_1,t)$ now contains the \textit{mean} self-consistent fields averaged over $f_1$, instead of the exact unaveraged fields~\cite{Klim_Statfizika}. Such approximation is called the Hartree's mean-field approximation~\cite{Hartree_1928}.

\subsection{Kinetic equation for one-particle quantum distribution function}
We consider a gas of spinless charged quantum particles with charge $e$ and mass $m$ (e.g., electrons with neglected spin), whose net charge is compensated by a background of heavy immobile particles (e.g., ions) with density $n_0$. Neglecting two-particle correlations due to exchange interactions and collisions between particles (see the discussion of collisionless mean-free approximation in Introduction), the quantum kinetic equation for the one-particle Wigner function $f_1$ of plasma particles is obtained from (\ref{eq:f_N_kinetic}) for $N=1$ and has the form (in what follows, we drop the index $1$ of the one-particle Wigner function, for brevity)
\begin{eqnarray}
\frac{\partial f(\mathbf{r},\mathbf{P},t)}{\partial t} &=& \frac{1}{(2\pi)^{6}}\frac{i}{\hbar}\int\ldots\int d\vec{\lambda}d\mathbf{k}d\vec{\eta}d\mathbf{q}\ {\rm e}^{i\left[\vec{\lambda}\cdot(\vec{\eta}-\mathbf{P})+\mathbf{k}\cdot(\mathbf{q}-\mathbf{r})\right]}
f(\mathbf{q},\vec{\eta},t)
\nonumber\\
&&\times\left[\mathcal{H}\left(\mathbf{q}-\frac{1}{2}\hbar\vec{\lambda},\vec{\eta}+\frac{1}{2}\hbar\mathbf{k},t\right)
- \mathcal{H}\left(\mathbf{q}+\frac{1}{2}\hbar\vec{\lambda},\vec{\eta}-\frac{1}{2}\hbar\mathbf{k},t\right)\right],
\label{eq:f_1_kinetic} \\
\mathcal{H}(\mathbf{r,P},t)&=&\frac{1}{2m}\left(\mathbf{P}-\frac{e}{c}\mathbf{A}(\mathbf{r},t)\right)^2 + e\phi(\mathbf{r},t), \label{eq:Hamiltonian}
\end{eqnarray}
where $\phi(\mathbf{r},t)$ and $\mathbf{A}(\mathbf{r},t)$ are respectively scalar and vector potentials of the self-consistent electromagnetic field. \textcolor{red}{Equations for $\phi$ and $\mathbf{A}$ are obtained from the Maxwell's equations, and their form depends on the chosen gauge condition.} For example, for the Coulomb gauge $\nabla\cdot\mathbf{A}=0$, we have the following coupled equations for $\phi$ and $\mathbf{A}$:
\begin{eqnarray}
- \nabla^2\phi &=& 4\pi \left[e n_e(\mathbf{r},t) - e n_0\right], \label{eq:phi} \\
\frac{1}{c^2}\frac{\partial^2\mathbf{A}}{\partial t^2} - \nabla^2\mathbf{A} +\frac{1}{c}\frac{\partial\nabla\phi}{\partial t}&=& \frac{4\pi}{c}\mathbf{j}_e(\mathbf{r},t), \label{eq:A}
\end{eqnarray}
\textcolor{red}{where the electron charge and current densities are defined as the zeroth- and first-order moments of the Wigner function:
\begin{eqnarray}
e n_e(\mathbf{r},t)=e\int{f(\mathbf{r,P},t)d\mathbf{P}}, \label{eq:ene}\\
\mathbf{j}_e(\mathbf{r},t) = \frac{e}{m}\int{\mathbf{p}f(\mathbf{r,P},t)d\mathbf{P}}. \label{eq:je}
\end{eqnarray}}
Here $\mathbf{P}=\mathbf{p}+(e/c) \mathbf{A}(\mathbf{r},t)$ is the canonical momentum, and $\mathbf{p}=m\mathbf{v}$ is the kinetic momentum of a quantum particle (we use CGS units). \textcolor{red}{Equations (\ref{eq:f_1_kinetic})--(\ref{eq:je}) form a closed set of equations describing electrodynamics of quantum plasma with electromagnetic interaction of particles.}

\subsection{Reduction to the Vlasov-Boltzmann-type kinetic equation}
Eq.~(\ref{eq:f_1_kinetic}) with (\ref{eq:Hamiltonian}) can be cast in the form of the Vlasov-Boltzmann-type kinetic equation in which all the explicit quantum terms are grouped separately, by the following steps:
\begin{itemize}
\item{Change of variables in (\ref{eq:f_1_kinetic})--(\ref{eq:Hamiltonian}) from $\mathbf{r,P},t$ to $\mathbf{r,p},t$, noting that
\begin{equation}
\frac{\partial f(\mathbf{r,P},t)}{\partial t} = \frac{\partial f(\mathbf{r,p},t)}{\partial t} - \frac{e}{c}\frac{\partial\mathbf{A}(\mathbf{r},t)}{\partial t}\cdot\frac{\partial f(\mathbf{r,p},t)}{\partial\mathbf{p}}, \label{eq:P->p}
\end{equation}}
\item{The next step is to change the integration variable $\vec\eta$ to $\vec\xi=\vec\eta-(e/c)\mathbf{A}$, and perform the integration over $\mathbf{q}$ and $\mathbf{k}$ in the right hand side of (\ref{eq:f_1_kinetic}). One thus obtains the following intermediate equation for $f(\mathbf{r,p},t)$:
\begin{eqnarray}
\frac{\partial f}{\partial t} + \left(\frac{\mathbf{p}}{m}+ \frac{e}{mc}\mathbf{A}\right)\cdot\frac{\partial f}{\partial\mathbf{r}} - \frac{e}{c}\frac{\partial\mathbf{A}}{\partial t}\cdot\frac{\partial f}{\partial\mathbf{p}} - \frac{e}{mc}\left(\mathbf{p}+\frac{e}{c}\mathbf{A}\right)\cdot\frac{\partial}{\partial\mathbf{r}}\left[\mathbf{A}\cdot\frac{\partial f}{\partial\mathbf{p}}\right] \nonumber \\
=\frac{1}{(2\pi)^3}\frac{1}{m}\iint d\vec\lambda d\vec\xi\ {\rm e}^{i\vec\lambda(\vec\xi-\mathbf{p})} \left\{-iem f(\mathbf{r},\vec\xi,t)\hat{D}_{\mathbf{r},\vec\lambda}[\phi(\mathbf{r},t)] \right. \nonumber \\
\left. + \frac{ie}{c}f(\mathbf{r},\vec\xi,t)\left(\vec\lambda+\frac{e}{c}\left[2\mathbf{A}(\mathbf{r},t) - \mathbf{A}(\mathbf{r}_{-},t)-\mathbf{A}(\mathbf{r}_{+},t)\right]\right)\cdot\hat{D}_{\mathbf{r},\vec\lambda}[\mathbf{A}(\mathbf{r},t)]\right. \nonumber \\
\left.+\frac{e}{2c}\left(\nabla f(\mathbf{r},\vec\xi,t) + f(\mathbf{r},\vec\xi,t)\nabla\right)\cdot\left[\mathbf{A}(\mathbf{r}_{-},t)+\mathbf{A}(\mathbf{r}_{+},t)\right] \right. \nonumber \\
\left. -\frac{e^2}{2c^2}\left[\left([\mathbf{A}(\mathbf{r}_{-},t)+\mathbf{A}(\mathbf{r}_{+},t)]\cdot\nabla\right)\mathbf{A}(\mathbf{r},t)\right]\cdot\frac{\partial f(\mathbf{r},\vec\xi,t)}{\partial\vec\xi} \right\}, \label{eq:app:intermediate}
\end{eqnarray}
where the ``spatial difference'' operator $\hat{D}_{\mathbf{r},\vec\lambda}$ is defined as $\hat{D}_{\mathbf{r},\vec\lambda}[g(\mathbf{r},t)]=[{g(\mathbf{r}_{+},t)-g(\mathbf{r}_{-},t)}]/{\hbar}$, with $\mathbf{r}_{\pm}=\mathbf{r}\pm\hbar\vec\lambda/2$ and $g(\mathbf{r},t)$ being a scalar or a vector function.}
\item{To Eq.~(\ref{eq:app:intermediate}), we add the following identities:
\begin{eqnarray}
\frac{1}{(2\pi)^3}\frac{1}{m}\iint d\vec\lambda d\vec\xi\ {\rm e}^{i\vec\lambda(\vec\xi-\mathbf{p})} \left\{-\frac{ie}{c}f(\mathbf{r},\vec\xi,t)(\vec\lambda\cdot\nabla)(\vec\xi\cdot\mathbf{A})\right\} \nonumber \\
= \frac{e}{mc}\left[\mathbf{p}\times\nabla\times\mathbf{A} + (\mathbf{p}\cdot\nabla)\mathbf{A}\right]\cdot\frac{\partial f(\mathbf{r,p},t)}{\partial\mathbf{p}} + \frac{e}{mc}(\nabla\cdot\mathbf{A})f(\mathbf{r,p},t),
\end{eqnarray}
\begin{eqnarray}
\frac{1}{(2\pi)^3}\frac{1}{m}\iint d\vec\lambda d\vec\xi\ {\rm e}^{i\vec\lambda(\vec\xi-\mathbf{p})} \left\{iem (\tau\cdot\nabla\phi)f(\mathbf{r},\vec\xi,t)\right\} = -e\nabla\phi\cdot\frac{\partial f}{\partial\mathbf{p}},
\end{eqnarray}
\begin{eqnarray}
-\frac{1}{(2\pi)^3}\frac{1}{m}\iint d\vec\lambda d\vec\xi\ {\rm e}^{i\vec\lambda(\vec\xi-\mathbf{p})} \left\{\frac{e}{c}\left(\nabla f(\mathbf{r},\vec\xi,t) + f(\mathbf{r},\vec\xi,t)\nabla\right)\cdot\mathbf{A}\right\} \nonumber \\
= -\frac{e}{mc}\mathbf{A}\cdot\nabla f(\mathbf{r,p},t) - \frac{e}{mc}(\nabla\cdot\mathbf{A})f(\mathbf{r,p},t),
\end{eqnarray}
\begin{eqnarray}
\frac{1}{(2\pi)^3}\frac{1}{m}\iint d\vec\lambda d\vec\xi\ {\rm e}^{i\vec\lambda(\vec\xi-\mathbf{p})} \left\{-\frac{ie^2}{c^2}f(\mathbf{r},\vec\xi,t)\vec\lambda\cdot\left[(\mathbf{A}\cdot\nabla)\mathbf{A}\right]\right\} \nonumber \\
= \frac{e^2}{mc^2}\left[(\mathbf{A}\cdot\nabla)\mathbf{A}\right]\cdot\frac{\partial f(\mathbf{r,p},t)}{\partial\mathbf{p}}.
\end{eqnarray}
}
\item{Collecting terms, and noting that
\[
\frac{\partial f(\mathbf{r},\vec\xi,t)}{\partial\vec\xi}\ {\rm e}^{i\vec\lambda\cdot(\vec\xi-\mathbf{p})} = \frac{\partial}{\partial\vec\xi}\left[f(\mathbf{r},\vec\xi,t)\ {\rm e}^{i\vec\lambda\cdot(\vec\xi-\mathbf{p})}\right] - i\vec\lambda f(\mathbf{r},\vec\xi,t)\ {\rm e}^{i\vec\lambda\cdot(\vec\xi-\mathbf{p})},
\]
we finally arrive at the desired result shown below.}
\end{itemize}

\subsection{Result: Vlasov-Boltzmann-type kinetic equation with ``quantum interference integral''}
The above procedure yields the following resulting form of the kinetic equation (\ref{eq:f_1_kinetic}) for the quantum distribution function $f(\mathbf{r,p},t)$:
\begin{eqnarray}
\frac{\partial f}{\partial t} + \frac{\mathbf{p}}{m}\cdot\frac{\partial f}{\partial\mathbf{r}} + e\left[\mathbf{E} + \frac{\mathbf{p}\times\mathbf{B}}{mc}\right]\cdot\frac{\partial f}{\partial\mathbf{p}} = I_q(\mathbf{r,p},t), \label{eq:f1_VB}
\end{eqnarray}
where $\mathbf{E}=-\nabla\phi - (1/c)\partial\mathbf{A}/\partial t$ and $\mathbf{B} = \nabla\times\mathbf{A}$ are the electric and magnetic field strengths, respectively.
The left hand side of (\ref{eq:f1_VB}) formally coincides with that of the classical Vlasov-Boltzmann equation (but for the quantum distribution function, which by itself contains quantum effects), and the right hand side respresents the ``quantum interference integral'' $I_q(\mathbf{r,p},t)$ defined as
\begin{eqnarray}
I_q(\mathbf{r,p},t)&=&\frac{1}{(2\pi)^3}\iint d\vec{\lambda}d\vec{\xi}\exp[i\vec{\lambda}\cdot(\vec{\xi}-\mathbf{p})] \nonumber \\
&\times&\left\{{ie}f(\mathbf{r},\vec{\xi},t)\left(\vec\lambda\cdot\frac{\partial\phi(\mathbf{r},t)}{\partial\mathbf{r}}-\hat{D}_{\mathbf{r},\vec\lambda}[\phi(\mathbf{r},t)]\right)\right. \nonumber \\ &&\left. -\frac{ie}{mc}f(\mathbf{r},\vec{\xi},t)\left(\vec\lambda\cdot\frac{\partial[\vec\xi\cdot\mathbf{A}(\mathbf{r},t)]}{\partial\mathbf{r}}-\hat{D}_{\mathbf{r},\vec\lambda}[\vec\xi\cdot\mathbf{A}(\mathbf{r},t)]\right) \right. \nonumber \\
&&\left. -\frac{e}{2mc}\left[\left(\frac{\partial f(\mathbf{r},\vec\xi,t)}{\partial\mathbf{r}} + f(\mathbf{r},\vec\xi,t)\frac{\partial}{\partial\mathbf{r}}\right) +\frac{ie}{c}f(\mathbf{r},\vec\xi,t) \left(\frac{\partial(\vec\lambda\cdot\mathbf{A}(\mathbf{r},t))}{\partial\mathbf{r}}-\hat{D}_{\mathbf{r},\vec\lambda}[\mathbf{A}(\mathbf{r},t)]\right)\right]\right. \nonumber \\
&&\left.\cdot\left[2\mathbf{A}(\mathbf{r},t) - \mathbf{A}(\mathbf{r}_{-},t)-\mathbf{A}(\mathbf{r}_{+},t)\right] \right\}. \label{eq:I_q}
\end{eqnarray}
The ``quantum interference integral'' $I_q$ manifests the effect of quantum interference of overlapping particle wave functions. It vanishes in the classical limit, which can be seen by formally taking the limit $\hbar\rightarrow 0$ in (\ref{eq:I_q}). In this case, the kinetic equation (\ref{eq:f1_VB}) reduces to the classical Vlasov equation for the classical distribution function. \textcolor{red}{Note that the kinetic equation (\ref{eq:f1_VB}) and the ``quantum interference integral'' (\ref{eq:I_q}) are derived without assuming any gauge fixing for the electromagnetic field potentials, and thus can be used in their present form for any gauge.}

\section{Use of the model}
\subsection{Numerical simulations of nonlinear electromagnetic processes in quantum plasmas}
Quantum kinetic equation (\ref{eq:f1_VB}) is presented in a convenient form that allows to incorporate quantum effects into existing classical electromagnetic Vlasov (Vlasov-Maxwell) codes~\cite{mangeney_etal_2002,eliasson_2003,shoucri_2008}, that can then be used for numerical simulation of nonlinear electrodynamic processes (e.g., modulational interaction of electrostatic and electromagnetic waves) in quantum plasma structures. Indeed, the left hand side of (\ref{eq:f1_VB}) has a form of classical Vlasov equation for the quantum distribution function $f$ (which is however not equivalent to the classical distribution function), while all the quantum interference effects due to overlapping of plasma particle wave functions are contained in the right hand side of (\ref{eq:f1_VB}). The six-dimensional ``quantum interference integral'' $I_q(\mathbf{r,p},t)$ (\ref{eq:I_q}) conserves the spatial number density of particles:
\begin{equation}
\int{I_q(\mathbf{r,p},t) d\mathbf{p}}=0, \label{eq:conservation}
\end{equation}
which can readily be seen by integrating (\ref{eq:I_q}) over $\mathbf{p}$. In the classical limit $\hbar\rightarrow0$, $I_q(\mathbf{r,p},t)$ tends to zero as $\hbar^2$:
\begin{eqnarray}
I_q(\mathbf{r,p},t) &=& \frac{\hbar^2}{24}\frac{1}{(2\pi)^3}\iint d\vec\lambda d\vec\xi\ {\rm e}^{i\vec\lambda\cdot(\vec\xi-\mathbf{p})} \nonumber \\
&\times&\left\{ -ie f(\mathbf{r},\vec\xi,t)(\vec\lambda\cdot\nabla)^3\phi + \frac{ie}{mc}f(\mathbf{r},\vec\xi,t)\left(\vec\xi\cdot\left[(\vec\lambda\cdot\nabla)^3 \mathbf{A}\right]\right) \right. \nonumber\\
&-& \left. \frac{3e}{mc}\left(\nabla f(\mathbf{r},\vec\xi,t) + f(\mathbf{r},\vec\xi,t)\left[\nabla + \frac{ie}{c}\vec\lambda\times(\nabla\times\mathbf{A})\right]\right)\cdot\left[(\vec\lambda\cdot\nabla)^2\mathbf{A}\right] \right\}. \label{eq:I_semiclassical}
\end{eqnarray}

To include quantum effects into an existing classical Vlasov-Maxwell numerical simulation code, only the following two modifications of the code are needed: 
\subsubsection{Initial equilibrium distribution \label{sec:f0}}
Introduction of proper initial (equilibrium $f_0(\mathbf{r,p})$ and initial perturbation $\tilde{f}(\mathbf{r,p},0)$) and boundary conditions for the quantum distribution function (Wigner function) $f(\mathbf{r,p},t)$~\cite{Manfredi_Howto}. Generally, these conditions differ from those for the classical distribution function, since the Wigner function is not a positively defined quantity and hence is not equivalent to particle density in coordinate-momentum phase space~\cite{Tatarskii_1983}. The equilibrium quantum distribution function should take into account the proper quantum statistics (occupation probabilities for particle states with given energies) for the considered type of quantum particles (e.g., the Fermi-Dirac statistics for electrons, which is not consistent with the approximation of spinless particles used in this paper, but still yields correct results, e.g., when modeling electrostatic phenomena in electron plasma). \textcolor{red}{We note that the equilibrium distribution function could in principle be defined by inclusion into the kinetic model of the collision integral for quantum particles $St[f]$ that accounts for both Coulomb and exchange interaction of particles, thus automatically ensuring Fermi or Bose statistics for fermions or bosons, respectively. With the proper $St[f]$ included in the model, the equilibrium distribution function $f_0(\mathbf{r,p})$ of quantum plasma would be defined self-consistently from the equation
\begin{equation}
\frac{\mathbf{p}}{m}\cdot\frac{\partial f_0}{\partial\mathbf{r}} + e\left[\mathbf{E}_0 + \frac{\mathbf{p}\times\mathbf{B}_0}{mc}\right]\cdot\frac{\partial f_0}{\partial\mathbf{p}} - I_{q0}(\mathbf{r,p}) = St[f_0].  \label{eq:f0_wouldbe}
\end{equation}
However, inclusion of the proper quantum collision integral $St[f]$ greatly complicates the kinetic model. Luckily, for many quantum plasmas (e.g., conduction electrons in metals) the characteristic collision frequency of particles is many orders of magnitude smaller than the characteristic frequency of plasma dynamics (plasma frequency)~\cite{Manfredi_Haas_PRB_2001}, so that collisions are not important for plasma dynamics (but are important for its equilibrium). In such cases, one can still use the collisionless kinetic model for the plasma dynamical response (thus avoiding the complications of introducing the collision integral), while \textit{postulating} (rather than obtaining self-consistently from Eq.~(\ref{eq:f0_wouldbe})) the equilibrium distribution function $f_0$ that mimics the effect of the missing $St[f_0]$ on defining the equilibrium. The postulated $f_0$ should be constructed from the wave functions of stationary states of plasma particles in the ``equilibrium'' electromagnetic field defined by $\phi_0$ and $\mathbf{A}_0$, accounting for the occupational probabilities for the corresponding states; see, e.g., Ref.~\cite{Kelly_1964}.}
\subsubsection{``Quantum interference integral''} 
Introduction of $I_q(\mathbf{r,p},t)$ that acts like a source term in the quantum-modified ``Vlasov'' kinetic equation for the quantum distribution function, redistributing $f(\mathbf{r,p},t)$ in coordinate-momentum phase space, while preserving the spatial number density of particles according to (\ref{eq:conservation}). Note that the dimensionality of the integral in $I_q(\mathbf{r,p},t)$ is significantly reduced by using symmetries of a chosen ad hoc problem, from the general six-dimensional integral down to a two-dimensional integral for one-dimensional problems, and to a four-dimensional integral for two-dimensional problems. Thus the evaluation of $I_q(\mathbf{r,p},t)$ is not expected to be prohibitively computationally expensive, especially for one-dimensional problems. In regimes when quantum effects are small but finite (i.e., when $\hbar$ is small compared with all plasma parameter combinations with dimensionality of $\hbar$; e.g., $e m^{1/2} n^{-1/6}$), approximation (\ref{eq:I_semiclassical}) for $I_q(\mathbf{r,p},t)$ should be used instead, which can be further simplified by using symmetries of the problem (e.g., in one-dimensional case the integration in (\ref{eq:I_semiclassical}) can be done analytically to the end).

These modifications are convenient since they do not affect the main parts of the classical Vlasov code: advection and acceleration of $f(\mathbf{r,p})$ in coordinate-momentum phase space, evaluation of charge and current densities, and solution of equations (\ref{eq:phi})--(\ref{eq:A}) for the scalar and vector potentials of the self-consistent electromagnetic field, -- all remain unchanged, owing to the formal similarity of the 
quantum kinetic equation (\ref{eq:f1_VB}) to the classical Vlasov-Boltzmann equation. Thus the suggested modification of an existing classical electromagnetic Vlasov code~\cite{mangeney_etal_2002,eliasson_2003,shoucri_2008}, instead of developing an electromagnetic analogue of Wigner-Poisson code~\cite{Suh_etal_1991} that would numerically solve Eq.~(\ref{eq:f_1_kinetic}) with computationally prohibitive integral on the right hand side, offers, in our view, a much quicker and more straightforward path to simulating linear and nonlinear electrostatic and electromagnetic phenomena in nonrelativistic quantum plasmas. 

\subsection{Linear response of uniform isotropic quantum plasma \label{sec:linear_response}}
To obtain the linear dielectric response of a quantum plasma, we consider a small perturbation $\tilde{f}(\mathbf{r,p},t)$ to equilibrium characterized by quantum distribution function $f_0(\mathbf{r,p})$ formed in the presence of stationary electromagnetic field characterized by ``equilibrium'' potentials $\phi_0$ and $\mathbf{A}_0$. (We note again that $f_0$ can not be obtained self-consistently from the kinetic model (\ref{eq:f1_VB}), but rather needs to be postulated (constructed from stationary states of plasma particles in the ``equilibrium'' field defined by $\phi_0$ and $\mathbf{A}_0$), as discussed in Sec.~\ref{sec:f0}.)

In case of a uniform isotropic equilibrium, we have $\phi_0=\mathbf{A}_0=0$, $f_0(\mathbf{r,p})=f_0(p)$. Substituting $f(\mathbf{r,p},t)=f_0(p) + \tilde{f}(\mathbf{r,p},t)$ with $|\tilde{f}|\ll f_0$ into (\ref{eq:f1_VB}) and linearizing, one obtains, assuming $\nabla f_0=0$ (uniform equilibrium) and $\nabla\cdot\mathbf{A}=0$ (Coulomb gauge), the following linear equation for $\tilde{f}(\mathbf{r,p},t)$:
\begin{eqnarray}
\frac{\partial \tilde{f}}{\partial t} + \frac{\mathbf{p}}{m}\cdot\frac{\partial \tilde{f}}{\partial\mathbf{r}} + e\left[-\nabla\phi - \frac{1}{c}\frac{\partial\mathbf{A}}{\partial t} + \frac{\mathbf{p}\times\nabla\times\mathbf{A}}{mc}\right]\cdot\frac{\partial f_0}{\partial\mathbf{p}} = \frac{ie}{(2\pi)^3}\iint d\vec{\lambda}d\vec{\xi}{\ \rm e}^{i\vec{\lambda}\cdot(\vec{\xi}-\mathbf{p})}f_0(\xi) \nonumber \\
\times\left\{\left(\vec\lambda\cdot\frac{\partial\phi(\mathbf{r},t)}{\partial\mathbf{r}}-\hat{D}_{\mathbf{r},\vec\lambda}[\phi(\mathbf{r},t)]\right) -\frac{1}{mc}\left(\vec\lambda\cdot\frac{\partial[\vec\xi\cdot\mathbf{A}(\mathbf{r},t)]}{\partial\mathbf{r}}-\hat{D}_{\mathbf{r},\vec\lambda}[\vec\xi\cdot\mathbf{A}(\mathbf{r},t)]\right) \right\}. \label{eq:f1_lin}
\end{eqnarray}
Fourier transforming (\ref{eq:f1_lin}), integrating over $\vec\lambda$ and $\vec\xi$, and solving for $\tilde{f}$, we obtain $\tilde{f}(t,\mathbf{r,p}) = (2\pi)^{-4} \iint{d\omega d\mathbf{k} \tilde{f}_{\omega,\mathbf{k}}(\mathbf{p}) \exp[-i(\omega t - \mathbf{k\cdot r})]}$, with
\begin{eqnarray}
\tilde{f}_{\omega,\mathbf{k}}(\mathbf{p}) = \left(-e\phi_{\omega,\mathbf{k}}+ \frac{e}{mc}\left(\mathbf{p\cdot\mathbf{A}_{\omega,\mathbf{k}}}\right)\right)\frac{\hat{D}_{\mathbf{p,k}}[f_0]}{\omega - \mathbf{k\cdot p}/m} + \frac{e}{mc}\left(\mathbf{p\cdot\mathbf{A}_{\omega,\mathbf{k}}}\right) \frac{\partial f_0}{\partial\epsilon},
\end{eqnarray}
where $\phi_{\omega,\mathbf{k}}$ and $\mathbf{A}_{\omega,\mathbf{k}}$ are the Fourier transforms of $\phi(\mathbf{r},t)$ and $\mathbf{A}(\mathbf{r},t)$, respectively, $\epsilon=p^2/2m$ is kinetic energy of plasma particles, and the ``momentum difference'' operator $\hat{D}_{\mathbf{p,k}}$ is defined as $\hat{D}_{\mathbf{p,k}}[f_0(\mathbf{p})] = [f_0(\mathbf{p}+\hbar\mathbf{k}/2)-f_0(\mathbf{p}-\hbar\mathbf{k}/2)]/\hbar$. Finally, substituting the obtained $\tilde{f}_{\omega,\mathbf{k}}(\mathbf{p})$ into the Fourier transformed Eq.~(\ref{eq:A}), after some straightforward manipulations one obtains the linear dielectric permittivity tensor $\varepsilon_{ij}(\omega,\mathbf{k})$ of uniform isotropic quantum plasma in the form
\[
\varepsilon_{ij}(\omega,\mathbf{k}) = \varepsilon^l(\omega,\mathbf{k}) \frac{k_ik_j}{k^2} + \varepsilon^{tr}(\omega,\mathbf{k}) \left(\delta_{ij} - \frac{k_ik_j}{k^2}\right),
\]
with the longitudinal and transverse permittivities given by
\begin{eqnarray}
\varepsilon^l(\omega,\mathbf{k}) &=& 1 + \frac{4\pi e^2}{k^2}\int{d\mathbf{p} \frac{\hat{D}_{\mathbf{p,k}}(f_0)}{\omega - \mathbf{k\cdot p}/m}}, \label{eq:eps_l} \\
\varepsilon^{tr}(\omega,\mathbf{k}) &=& 1 -\frac{\omega_p^2}{\omega^2} + \frac{2\pi e^2}{m^2\omega^2}\int{d\mathbf{p}\ p_\perp^2\frac{\hat{D}_{\mathbf{p,k}}(f_0)}{\omega-\mathbf{k\cdot p}/m}}, \label{eq:eps_tr}
\end{eqnarray}
where $p_\perp$ is the absolute value of the component of $\mathbf{p}$ perpendicular to $\mathbf{k}$. 

Note that in (\ref{eq:eps_l})--(\ref{eq:eps_tr}) the integration over $p_\parallel$ (a component of $\mathbf{p}$ along $\mathbf{k}$) is undefined due to a singularity (simple pole) at $p_\parallel=m\omega/k$, for real $\omega,k$. This difficulty is avoided by requiring that all perturbations obey the causality principle, i.e., appear at some initial time $t_0$. For such perturbations, their temporal Fourier transforms become one-sided (e.g., $\phi_\omega=\int_0^\infty{\phi(t')\exp(i\omega t') dt'}$, with $t'=t-t_0$), and are defined for complex $\omega$ with ${\rm Im}(\omega)>0$. To extend their definition to real $\omega$, one needs to perform their analytical continuation onto the real axis ${\rm Im}(\omega)=0$ of complex $\omega$ plane, which can be done by taking the limit ${\rm Im}(\omega)\rightarrow 0+$ in the corresponding functions of complex $\omega$. This procedure, known as Landau's rule, defines how the contour of integration over $p_\parallel$ in (\ref{eq:eps_l})--(\ref{eq:eps_tr}) should be deformed to avoid the singularity at $p_\parallel=m\omega/k$, and is equivalent to replacing $\omega$ with ${\rm Re}(\omega)+io$ and taking the limit $o\rightarrow 0+$, followed by integrating along real $p_\parallel$. The resulting dielectric permittivity can be complex, its imaginary part leading to Landau damping~\cite{Landau_1946}.

The obtained linear longitudinal and transverse dielectric permittivities (\ref{eq:eps_l})--(\ref{eq:eps_tr}) match with those previously obtained in the literature for uniform isotropic weakly coupled quantum plasma of spinless particles~\cite{Klim_Silin_UFN_1960,Kuz'menkov_Maksimov_2002,Eliasson_Shukla_JPP_2009}, and define, e.g., dispersion properties of small-amplitude electrostatic and electromagnetic oscillations in such plasmas.

\subsection{Applicability \label{sec:applicability}}
The quantum kinetic equation (\ref{eq:f1_VB}) is obtained for weakly coupled nonrelativistic quantum plasma. The assumption of weak coupling is justified when the coupling parameter $\Gamma=U_{\rm int}/\epsilon_{\rm kin}$ is small, which in case of plasma consisting of degenerate electrons with density $n$ and Fermi energy $\epsilon_F=(\hbar^2/2m)(3\pi^2 n)^{2/3}$ implies $\Gamma\sim e^2 n^{1/3}/\epsilon_F \ll 1$, which is equivalent to 
\begin{equation}
e^2/\hbar v_F\ll 1,	\label{eq:Gamma<<1}
\end{equation}
where $v_F=\sqrt{2 \epsilon_F/m}$ is the electron Fermi velocity. (Note that, even if the spin of plasma electrons is taken into account, the corresponding effect of electron correlation due to their exchange interaction is negligibly small when (\ref{eq:Gamma<<1}) is satisfied~\cite{Klim_Silin_UFN_1960}.) On the other hand, the assumption of nonrelativistic plasma, $v\ll c$, where $v$ is the characteristic velocity of plasma particles, and $c$ is the speed of light, implies in the case of degenerate electron plasma (for which $v\sim v_F$) that
\begin{equation}
e^2/\hbar v_F\gg e^2/\hbar c \approx 1/137.	\label{eq:v_F<<c}
\end{equation}

If the condition (\ref{eq:Gamma<<1}) is violated (e.g., for electrons in metals), the effect of particle correlations (collisions and exchange interactions) becomes important, and such plasmas should be treated as a quantum liquid; however, the present model (\ref{eq:f1_VB}) may still yield qualitatively correct results for such plasmas~\cite{Klim_Silin_UFN_1960}. In plasmas where condition (\ref{eq:v_F<<c}) is violated (e.g., in dense plasmas subject to ultrarelativistic laser radiation, or in the matter of white dwarfs), relativistic effects become important. Both conditions (\ref{eq:Gamma<<1}) and (\ref{eq:v_F<<c}) may be met simultaneously in some semiconductors. 

The model presented here ignores the spin of plasma particles, which makes it formally valid only for quantum plasmas consisting of spinless particles, with conditions (\ref{eq:Gamma<<1}) and (\ref{eq:v_F<<c}) simultaneously met. On the other hand, our kinetic model is more general than any quantum fluid model, as it accounts for essentially kinetic effects which the fluid models ignore completely, and thus it should be used in cases where such effects are expected to be important. Generalizing our kinetic model to include either (or, ideally, all) of the ignored effects (spin~\cite{Klim_Silin_UFN_1960,Brodin_etal_2011}, plasma particle correlations~\cite{Klim_Silin_UFN_1960}, and relativistic effects) would be a natural continuation of this work.

\textcolor{red}{\subsection{On gauge invariance of the model}
The Wigner function $f(\mathbf{r,p},t)$ used in the kinetic model presented here explicitly depends on the potentials $\phi$ and $\mathbf{A}$, and thus it is not gauge-invariant~\cite{Haas_etal_NJP_2010,Serimaa_etal_PRA_1986}. Indeed, applying the gauge transformation of potentials $\mathbf{A'}=\mathbf{A}+\nabla g(\mathbf{r},t)$, $\phi'=\phi-(1/c)\partial g(\mathbf{r},t)/\partial t$ along with the simultaneous unitary transformation $\psi'=\psi\exp[(ie/\hbar c)g(\mathbf{r},t)]$ of particle wave functions (introduced in order to preserve the form of Schr\"{o}dinger equation with gauge-transformed potentials~\cite{L&L_3}), we have for the transformed Wigner function:
\begin{equation}
f'(\mathbf{r,p},t) = \frac{1}{(2\pi)^3}\int{d\vec\lambda \exp\left[-\frac{ie}{c}\left(\vec\lambda\cdot\nabla g - \hat{D}_{\mathbf{r},\vec\lambda}[g]\right)\right]{\rm e}^{-i\vec\lambda\cdot[\mathbf{p}+(e/c)\mathbf{A}]}\rho(\mathbf{r}-\hbar\vec\lambda/2,\mathbf{r}+\hbar\vec\lambda/2)}.
\end{equation}
In the classical limit, $\vec\lambda\cdot\nabla g - \hat{D}_{\mathbf{r},\vec\lambda}[g]$ vanishes, leading to $f'=f$, i.e., the Wigner function becomes the gauge-invariant classical distribution function, as expected. However, in general $f'(\mathbf{r,p},t)\neq f(\mathbf{r,p},t)$, i.e., the quantum distribution function $f(\mathbf{r,p},t)$ is \textit{not} gauge-invariant.} 

\textcolor{red}{The fact that $f(\mathbf{r,p},t)$ is gauge-dependent presents a problem of defining high-order (3rd and higher) moments of $f$ in a gauge-invariant way~\cite{Haas_etal_NJP_2010}. This problem may be important in the hydrodynamic model of quantum plasma, where the evolution of lower-order moments of $f$ relies on (and is affected by) the evolution of higher-order moments, which thus all have to be defined from $f(\mathbf{r,p},t)$ in a gauge-invariant way as the chain of hydrodynamic equations is derived from the kinetic model~\cite{Haas_etal_NJP_2010,Vlad_Tysh_UFN_2011}. However, in the \textit{kinetic} description of plasma electrodynamics developed here, the problem of gauge-dependent $f(\mathbf{r,p},t)$ does not present itself. Indeed, in order to have a complete description of self-consistent electromagnetic field in plasma, only the knowledge of charge and current density of plasma particles is required, as only these two quantities appear in the Maxwell's equations for the self-consistent electromagnetic field. In the kinetic description employing the gauge-dependent Wigner function $f(\mathbf{r,p},t)$, electron charge and current densities, defined by Eqs~(\ref{eq:ene})--(\ref{eq:je}) as the zeroth- and first-order moments of $f$, are \textit{gauge-invariant}, even though $f$ itself is not~\cite{Haas_etal_NJP_2010}; hence the plasma response to the electromagnetic field, defined from the gauge-dependent Wigner function, is gauge-invariant. Therefore, the kinetic equation (\ref{eq:f1_VB}) coupled with Maxwell's equations for $\phi$ and $\mathbf{A}$ (written for a chosen gauge), although using the gauge-dependent Wigner function $f(\mathbf{r,p},t)$, describes electrodynamics of quantum plasma in a gauge-invariant way. (We stress that Eqs~(\ref{eq:f1_VB})--(\ref{eq:I_q}) are derived without assuming any gauge fixing, and thus can be used in their present form for any gauge.)}

\textcolor{red}{In those cases where knowledge of third- or higher-order moments of $f(\mathbf{r,p},t)$ is required from a kinetic simulation, an alternative kinetic model employing a specially constructed gauge-invariant Wigner function may be of advantage~\cite{Haas_etal_NJP_2010}. Otherwise, for most practical purposes, the kinetic model developed here provides an adequate description of quantum plasma electrodynamics, subject to the applicability limits discussed in Sec.~\ref{sec:applicability}.}

\section{Conclusions}
In this paper, a kinetic theory of nonrelativistic quantum plasma with electromagnetic interaction of its particles is developed in the framework of the Hartree's mean-field approximation. The obtained quantum kinetic equation for the one-particle quantum distribution function (Wigner function) is cast in a form of the classical Vlasov equation with the additional quantum source term accounting for the quantum interference of overlapping plasma particle wave functions. This form of quantum kinetic equation suggests a straightforward modification of existing classical electromagnetic Vlasov-Maxwell codes, without affecting the core parts of their solvers, which opens a relatively quick path to simulating electromagnetic (as well as electrostatic) phenomena in various quantum plasma structures. Such modified Vlasov codes could also provide a useful basis for theoretical studies of quantum plasmas, as quantum and classical effects can be easily separated in such codes.

\acknowledgments{This work was supported by the Australian Research Council. R.K. acknowledges the receipt of a Professor Harry Messel Research Fellowship funded by the Science Foundation for Physics within the University of Sydney.}


\end{document}